\documentclass[12pt,superscriptaddress]{article}
\pdfoutput=1
\usepackage{jheppub}
\usepackage{epsfig}
\usepackage{amsmath,amssymb}
\usepackage{amsfonts}
\usepackage{mathrsfs}
\usepackage{units}

\newcommand{\beq}{\begin{equation}}
\newcommand{\eeq}[1]{\label{#1}\end{equation}}
\def\beqa{\begin{eqnarray}}
\def\eeqa#1{\label{#1}\end{eqnarray}}
\newcommand{\eeqn}{\end{equation}}
\newcommand{\CR}{\notag \\}
\newcommand{\leqn}[1]{\eqref{#1}}

\def\to{\rightarrow}
\def\s{s_\beta}
\def\c{c_\beta}
\def\r{x_t}
\def\mT{m_T}
\def\half{\frac{1}{2}}

\def\stacksymbols #1#2#3#4{\def\theguybelow{#2}
    \def\vp{\lower#3pt}
    \def\sp{\baselineskip0pt\lineskip#4pt}
    \mathrel{\mathpalette\intermediary#1}}

\def\intermediary#1#2{\vp\vbox{\sp
     \everycr={}\tabskip0pt
     \halign{$\mathsurround0pt#1\hfil##\hfil$\crcr#2\crcr
              \theguybelow\crcr}}}

\def\lsim{\stacksymbols{<}{\sim}{2.5}{.2}}
\begin{document}

\title{A Fermionic Top Partner: Naturalness \\ and the LHC}

\author[a]{Joshua Berger}
\author[b]{, Jay Hubisz}
\author[a]{, Maxim Perelstein}
\affiliation[a]{Laboratory for Elementary Particle Physics, Cornell University, Ithaca, NY 14853, USA}
\affiliation[b]{201 Physics Building, Syracuse University, Syracuse, NY 13244, USA}

\abstract{Naturalness demands that the quadratic divergence of the one-loop top contribution to the Higgs mass be cancelled at a scale below 1 TeV. This can be achieved by introducing a fermionic (spin-1/2) top partner, as in, for example, Little Higgs models. In this paper, we study the phenomenology of a simple model realizing this mechanism. We present the current bounds on the model from precision electroweak fits, flavor physics, and direct searches at the LHC. The lower bound on the top partner mass from precision electroweak data is approximately 500 GeV, while the LHC bound with 5 fb$^{-1}$ of data at $\sqrt{s}=7$ TeV is about 450 GeV. Given these bounds, the model can incorporate a 125 GeV Higgs with minimal fine-tuning of about 20\%. We conclude that natural electroweak symmetry breaking with a fermionic top partner remains a viable possibility. We also compute the Higgs decay rates into gauge bosons, and find that significant, potentially observable deviations from the Standard Model predictions may occur.}

\maketitle


\section{Introduction}

The Standard Model (SM) of particle physics postulates the existence of an elementary scalar field, the Higgs, which is responsible for electroweak symmetry breaking. Precision measurements of the properties of electroweak gauge bosons are consistent with this picture, and favor a light ($\sim 100$ GeV) Higgs boson. Recently, experiments at the Large Hadron Collider (LHC) reported preliminary evidence for a new particle with properties roughly consistent with the SM Higgs and a mass of about 125 GeV~\cite{LHC_Higgs}.

In the SM, the contribution of quantum loops to the Higgs mass term is quadratically divergent. To avoid fine-tuning, new physics beyond the SM must appear and cut off this divergence at a scale of order 1 TeV or below. Precision electroweak data favors models where the divergence is cancelled by loops of new weakly-coupled states; such cancellations can occur naturally as a consequence of underlying symmetries of the theory. 
What is the minimal set of new particles that must appear below 1 TeV to avoid fine-tuning? It is well known that the only SM contribution to the Higgs mass that must be modified at sub-TeV scales is the one-loop correction from the top sector. All other SM loops are numerically suppressed by either gauge or non-top Yukawa couplings, by extra loop factors, or both. As a result, the states responsible for cutting off these loops can lie above 1 TeV with no loss of naturalness. Thus, the sub-TeV particles that soften the divergence in the top loop, the  ``top partners," provide a uniquely well-motivated target for searches at the LHC, and it must be ensured that a comprehensive, careful search for such partners is conducted. 

The best-known mechanism for canceling the Higgs mass divergences is supersymmetry (SUSY). In SUSY models, the quadratic divergence in the SM top loop is cancelled by loops of scalar tops, or stops. Recently, a number of papers~\cite{NSUSY} emphasized the importance of stop searches at the LHC, and reinterpreted the published LHC results, based on the 1 fb$^{-1}$ integrated luminosity data set, in terms of bounds on stop masses. It was found that completely natural spectra are allowed so far. On the other hand, incorporating a 125 GeV Higgs in the Minimal SUSY Model (MSSM) 
does require significant fine-tuning, of order 1\% at best. (Fine-tuning can be reduced in non-minimal models~\cite{SUSY_Higgs_FT}.)

\begin{figure}[tb]
\begin{center}
\centerline {
\includegraphics{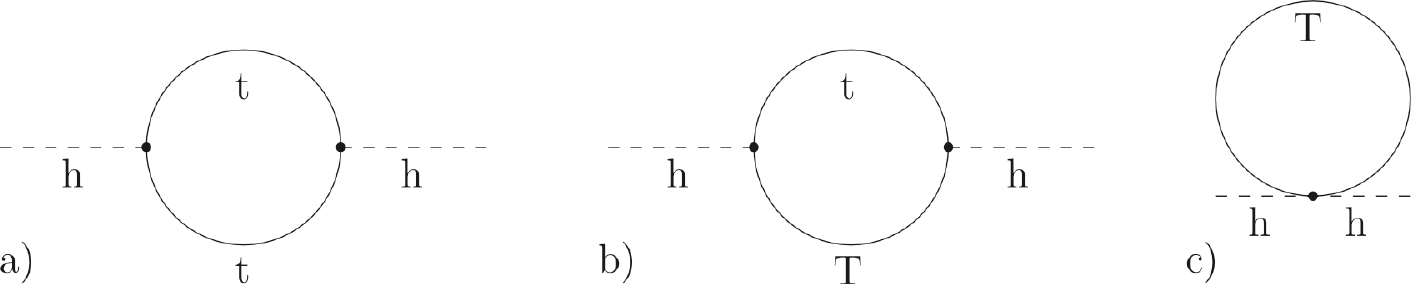}
}
\vspace{.2in}
\caption{One-loop Higgs mass renormalization in a model with a fermionic top partner, such as the Littlest Higgs.}
\label{fig:LHloops}
\end{center}
\end{figure}

However, SUSY is not the only option for canceling the quadratic divergence in the SM top loop. An alternative is to introduce a spin-$1/2$ top partner $T$, a Dirac fermion with mass $m_T$, which is an $SU(2)_L$ singlet, color triplet, and has electric charge $2/3$. In the Weyl basis, $T=(T_L,T_R)$. This field couples to the SM Higgs doublet $H$ via
\beq
{\cal L} = - \lambda_T  T_R^\dagger \tilde{H} Q_3 + \frac{\lambda_t^2 + \lambda_T^2}{2m_T} (H^\dagger H) T_L^\dagger T_R +~{\rm h.c.}\,,
\eeq{couplings}
where $Q_3$ is the SM third-generation left-handed quark doublet, $\lambda_t$ is the SM top Yukawa, $\lambda_T$ is a new dimensionless coupling constant, and  $\tilde{H} = (i\sigma_2 H)^\dagger$. The one-loop contribution to the Higgs mass in this model is shown in 
Fig.~\ref{fig:LHloops}; the quadratic divergences present in each of the three diagrams  cancel in the sum. Even though the structure of the couplings in Eq.~\leqn{couplings} looks completely ad hoc at first sight, it can emerge naturally if the Higgs is embedded as a pseudo-Nambu-Goldstone boson~\cite{HPNGB} of spontaneous global symmetry breaking at the TeV scale. The global symmetry must be broken explicitly to induce non-derivative Yukawa and gauge couplings of the Higgs; divergence cancellation is achieved if the explicit symmetry breaking terms obey the ``collective" condition, such as in Little Higgs models~\cite{LH,LHreviews}. (A similar mechanism is operative in the 5-dimensional composite Higgs models~\cite{5DHiggs}, where the role of the top partner is played by the Kaluza-Klein excitations of the top.) In this paper, we will
focus on a minimal model that incorporates the top Yukawa via collective symmetry breaking and explicitly realizes the structure of Eq.~\leqn{couplings}. We will present direct and indirect bounds on the model and discuss their implications for naturalness in light of the 125 GeV Higgs. We will also consider predictions for the deviations of the Higgs and top properties from the SM. 

Our model is basically identical to the top sector of the Littlest Higgs~\cite{LH}, and we will make use of many results derived in the context of that model. The original Littlest Higgs is severely constrained by precision electroweak data~\cite{LH_PEW}. The constraints come almost entirely from the extra gauge bosons of the model, whose masses are required to be above 2-3 TeV. In itself, this is not a problem for naturalness. However, the structure of the Littlest Higgs imposes a tight relation between the gauge boson and top partner masses, so that multi-TeV top partners are required, which in turn implies strong fine-tuning. This problem can be avoided by modifying the model, by introducing an additional symmetry (T-parity) to forbid tree-level corrections to precision electroweak observables~\cite{LHT}, by decoupling the top and gauge boson partner mass scales~\cite{Bestest}, or simply by slightly lowering the cutoff and getting rid of the extra gauge bosons altogether~\cite{interLH}. Thus, while the structure of the top sector is robust -- it is in effect fixed by the naturalness requirement -- the gauge and scalar sectors appear quite model-dependent, both in their structure and in the associated mass scale. Motivated by these considerations, we consider the top sector in isolation, and identify the predictions that are in a sense unavoidable once the cancellation mechanism in Fig.~\ref{fig:LHloops} is postulated. This approach is similar to the bottom-up attitude to SUSY phenomenology advocated in Refs.~\cite{NSUSY}. 

The rest of the paper is organized as follows. The minimal model for the fermionic top partner is presented in Section~\ref{sec:NLSM}. Section~\ref{sec:FT} discusses naturalness of electroweak symmetry breaking in this model, assuming a 125 GeV Higgs boson. Section~\ref{sec:bounds} summarizes existing experimental constraints on the model, divided in three groups: precision electroweak, flavor constraints, and direct searches at the LHC. Sections~\ref{sec:Higgs} and~\ref{sec:top} discuss the expected deviations of the Higgs and top properties, respectively, from the SM predictions. We summarize our findings and conclude in Section~\ref{sec:conc}. A number of useful formulas are collected in the Appendix. 

\section{Minimal Model for Fermionic Top Partner}
\label{sec:NLSM}

We begin with a non-linear sigma model describing spontaneous $SU(3)\rightarrow SU(2)$ global symmetry breaking by a fundamental vev. The sigma field is
\beq
V = \exp \left(\frac{i \pi_a t_a}{f}\right) \left( \begin{tabular}{c} 0\\0\\$f$ \end{tabular} \right) \,,
\eeq{sigma}
where $t_a$ are the broken generators ($a=1\ldots 5$), $\pi_a$ are the corresponding Goldstone bosons, and $f$ is the symmetry breaking scale (we assume $f\lsim 1$ TeV).  We identify the $SU(2)$ doublet of Goldstone bosons with the SM Higgs doublet $H$, and ignore the remaining one which plays no role in our analysis:
\beq
\pi_a t_a =  \left( \begin{tabular}{cc} 0 & $H$ \\ $H^\dagger$ & 0 \end{tabular} \right) \,.
\eeq{piH}
To generate a top Yukawa coupling without introducing one-loop quadratic divergences, we introduce an $SU(3)$ triplet of left-handed Weyl fermions, $\chi_L = (\sigma^2 Q, U)^T_L$, and two $SU(3)$ singlet right-handed Weyl fermions, $u_R$ and $U_R$. Here $Q_L = (t_L, b_L)$. These fields are coupled via~\cite{LH,PPP}
\beq
{\cal L} \,=\, -\lambda_1 u_R^\dagger  V^\dagger \chi_L \,-\, \lambda_2 f U_R^\dagger U_L \,+\,~{\rm h.c.}
\eeq{toy1}
Expanding the sigma field up to terms of order $1/f^2$ gives
\beq
{\cal L} \,=\,- f \left( \lambda_1 u_R + \lambda_2 U_R\right)^\dagger U_L \,-\, \lambda_1 u_R^\dagger  \tilde{H} Q_L \,+\, \frac{\lambda_1}{2f} (H^\dagger H) u_R^\dagger U_L +~{\rm h.c.}~+ \ldots   
\eeq{toy2}
where $\tilde{H} = (i\sigma_2 H)^\dagger$. The fermion mass eigenstates are 
\beqa
T_L = U_L, ~~~~T_R = \frac{\lambda_1 u_R + \lambda_2 U_R}{\sqrt{\lambda_1^2+\lambda_2^2}}\,,\CR
t_L = u_L,~~~~t_R = \frac{\lambda_2 u_R - \lambda_1 U_R}{\sqrt{\lambda_1^2+\lambda_2^2}}\,,
\eeqa{eigentops}
where we neglected the Higgs vev $v$, assuming $v\ll f$. (The mixing angles and masses with full $v$ dependence are given in Appendix A.) We identify $t=(t_L,t_R)$ with the SM top quark, $b_L$ with the SM left-handed bottom, and $T=(T_L,T_R)$ with the top partner, whose mass is 
\beq
m_T = \sqrt{\lambda_1^2+\lambda_2^2}\, f\,.
\eeq{Tmass}
The interaction terms in the mass eigenbasis become 
\beq
{\cal L}_{\rm int} \,=\, -\lambda_t t_R^\dagger  \tilde{H} Q_L \,-\, \lambda_T T_R^\dagger  \tilde{H} Q_L
\,+\, \frac{\lambda_1^2}{m_T} (H^\dagger H) T_R^\dagger T_L \,+\, \frac{\lambda_1\lambda_2}{2m_T} (H^\dagger H) t_R^\dagger T_L +~{\rm h.c.}~+\ldots 
\eeq{toy3}
where we defined
\beq
\lambda_t = \frac{\lambda_1\lambda_2}{\sqrt{\lambda_1^2+\lambda_2^2}}\,,~~~\lambda_T = \frac{\lambda_1^2}{\sqrt{\lambda_1^2+\lambda_2^2}}\,.
\eeq{lambdas}
The first term is simply the SM top Yukawa; the next two terms reproduce Eq.~\leqn{couplings}, ensuring the cancellation of the one-loop quadratic divergence (note that $\lambda_1^2=\lambda_t^2+\lambda_T^2$); while the last term does not contribute to the Higgs mass renormalization at one loop, and thus does not spoil the cancellation. The cancellation is also easy to understand in terms of symmetries of the model: the first term in~\leqn{toy1} preserves the global $SU(3)$, so that in the limit $\lambda_2\to 0$ the Higgs is an exact Goldstone boson and is therefore massless. On the other hand, the second term in~\leqn{toy1} breaks the $SU(3)$ explicitly, but it does not involve the Higgs at all, and so cannot generate the Higgs mass on its own. Thus, both couplings need to enter any diagram contributing to the Higgs mass renormalization, and at the one-loop level the diagrams involving both $\lambda$'s are at most logarithmically divergent. 

The Higgs can be given its usual SM gauge couplings by weakly gauging the $SU(2)\times U(1)$ subgroup of the $SU(3)$. 
As explained in the Introduction, we do not consider extended gauge sectors here: the gauge structure of our model is the same as SM. The new top-sector fields $U_L$, $U_R$ have the same gauge quantum numbers as the SM right-handed top, $({\bf 3}, {\bf 1})_{4/3}$.  

Non-linear sigma model interactions become strongly coupled at a scale $\Lambda \approx 4\pi f$, where another layer of new physics must occur. The effects of that physics on weak-scale observables can be parametrized by adding operators of mass dimension $>4$, suppressed by appropriate powers of $\Lambda$, to the lagrangian. The leading (dimension-6) operators are
\beq
{\cal L}_{\rm UV} = \frac{c_1}{(4\pi f)^2} \left( V^\dagger D_\mu V \right)^2 \,+\, \frac{g g^\prime c_2}{(4\pi f)^2} W^a_{\mu\nu} B^{\mu\nu}  (V^\dagger Q^a V)\,,
\eeq{d6ops}
where $D_\mu$ is the covariant derivative including the $SU(2)\times U(1)$ gauge fields; $W$ and $B$ are the $SU(2)$ and $U(1)$ field strength tensors, respectively; $c_1$ and $c_2$ are dimensionless coefficients, which are unknown but expected to be of order 1; and
\beq
Q^a = \left( \begin{tabular}{cc} $\sigma^a$ &  \\& $0$ \end{tabular} \right) \,.
\eeq{Qdef}
The two operators in Eq.~\leqn{d6ops} contribute to the $T$ and $S$ parameters, respectively, in precision electroweak fits (see Sec.~\ref{sec:pew}). We do not include operators involving the top quark, since they are not strongly constrained at present. 

\section{Higgs Mass and Naturalness}
\label{sec:FT}

An appealing feature of the class of models we're dealing with is a simple, rather predictive description of the electroweak symmetry breaking (EWSB). At tree level, the Higgs is a Goldstone boson and the Higgs mass parameter $\mu^2=0$. At one loop, the leading (log-divergent) contribution to the Higgs mass parameter from the diagrams in Fig.~\ref{fig:LHloops} is given by
\beq
\delta\mu^2 = - 3 \frac{\lambda_t^2 m_T^2}{8\pi^2}\log \frac{\Lambda^2}{m_T^2}\,.
\eeq{delta_mu} 
Naive dimensional analysis (NDA) suggests that this is the dominant contribution to the Higgs mass: two-loop quadratically divergent contributions are suppressed by a power of $\log (4\pi)^2 \approx 5$, while gauge boson loops (assuming that their quadratic divergences are canceled at a scale close to 1 TeV) are down by $(g/\lambda)^2$. Note that Eq.~\leqn{delta_mu}  automatically has the right (negative) sign to trigger EWSB.

If the LHC hint is correct and there is indeed a 125 GeV Higgs boson, then $\mu^2$ can be treated as known, since $m_h = \sqrt{2}|\mu|$. In our model, this essentially fixes the top partner mass, up to logarithmic dependence on $\Lambda$. For definiteness, we take $\Lambda=4\pi f$; to leading order in $v/f$,  
\beq
\Lambda\approx \frac{2\pi}{\lambda_t}\,m_T \sin2\alpha\,,
\eeq{Lam_def}
where $\alpha$ is the mixing angle in the right-handed top sector (at leading order at $v/f$, $\tan\alpha=\lambda_1/\lambda_2$). For example, for $\alpha=\pi/4$, we obtain 
\beq
m_T \approx 236~{\rm GeV}. 
\eeq{top_partner}
Unfortunately, the top partner at this mass is excluded by precision electroweak constraints, see Section~\ref{sec:bounds}. 
The mild $\alpha$ dependence does not change this conclusion if $\alpha$ is varied within a reasonable range.

\begin{figure}[tb]
\begin{center}
\centerline {
\includegraphics[width=4in]{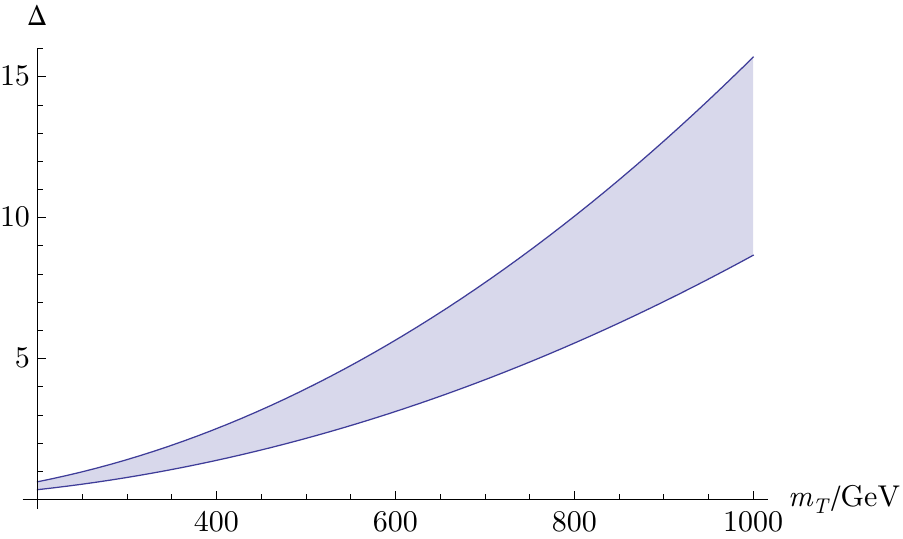}
}
\caption{Fine-tuning as a function of the top partner mass (in GeV). We fixed $m_h=125$ GeV. The band corresponds to varying the mixing angle $\alpha$ between $0.2$ and $1.1$.
}
\label{fig:FTplot}
\end{center}
\end{figure}

The only way to raise $m_T$ and salvage the model is to assume that the gauge-loop and/or two-loop contributions to $\mu^2$ are enhanced, and partially cancel the top-loop contribution.\footnote{In fact, Ref.~\cite{Grinstein} argued that the  two-loop contribution in the Littlest Higgs is enhanced compared to the NDA estimate, and estimated that it is of the same order as the logarithmically divergent one-loop contribution. Since the two-loop contribution is UV-dominated, its magnitude (and sign) cannot be determined without specifying a UV completion and performing a calculation in a UV-complete model.} This requires a certain degree of fine-tuning; we quantify it by defining
\beq
\Delta = \frac{|\delta\mu^2|}{\mu_{\rm obs}^2}\,,
\eeq{FT}   
where $\mu_{\rm obs} = m_h/\sqrt{2} \approx 88$ GeV. Required fine-tuning as a function of the top partner mass is shown in Fig.~\ref{fig:FTplot}, where the band corresponds to varying the mixing angle $\alpha$ between 0.2 and 1.1, corresponding roughly to the range where both $\lambda_1$ and $\lambda_2$ are perturbative. This plot should be kept in mind as we discuss the experimental constraints on the model below. 

A Higgs quartic coupling, $\lambda\approx 0.12$, is required to accommodate the Higgs vev $v=246$ GeV along with a 125 GeV mass. In our model, there is no tree-level quartic, but at one loop the quartic is generated by quadratically divergent terms in the Coleman-Weinberg potential~\cite{Coleman:1973jx,LH}.  In our minimal model, the quartic generated by the top-sector is in fact only logarithmically sensitive to the cutoff, and is thus expected to be small.  However, the contributions to global symmetry breaking due to gauging the SM $SU(2)_L \times U(1)_Y$ do generate quadratically divergent contributions to the quartic.  These diagrams are dominated by physics at the scale $\Lambda$, and hence cannot be computed without specifying a UV completion, but NDA estimates show that an ${\cal O}(g^2)$ quartic can be generated without tuning. 

\section{Experimental Constraints}
\label{sec:bounds}

The model in Eq.~\leqn{toy1} has three parameters: the symmetry breaking scale $f$ and two dimensionless couplings $\lambda_{1,2}$. One combination of the couplings has to be fixed to reproduce the known top Yukawa, leaving two independent parameters. In our discussion of experimental constraints, we will use the top partner mass $m_T$ and the rotation angle $\alpha$ between the gauge and mass eigenstates in the right-handed fermion sector. That is, $\alpha$ is defined by
\beq
t_R = \cos\alpha \,u_R - \sin\alpha\,u_R\,,~~~T_R = \sin\alpha \,u_R + \cos\alpha\,U_R\,.
\eeq{alpha_def}
The relation between ($m_T$, $\alpha$) and the Lagrangian parameters, to leading order in $v/f$, is given in Eqs.~\leqn{eigentops},~\leqn{Tmass}. In the analysis below, we will use generalizations of these formulas to all orders in $v/f$, see Appendix A. It is also worth noting that at order $v/f$, mixing between the left-handed fermion fields $u_L$ and $U_L$ is induced; the mixing angle $\beta$ is approximately given by
\beq
\sin\beta \approx \tan\alpha \,\frac{m_t}{m_T}\,.
\eeq{beta} 
Again, we will use the exact expression for this mixing angle, given in Appendix A. This mixing induces the off-diagonal vector boson couplings to fermions, $ZtT$ and $WbT$, which play a crucial role in the phenomenology of the model. Both couplings are proportional to $\sin\beta$. 

\subsection{Precision Electroweak Constraints}
\label{sec:pew}

The top partner $T$ does not induce tree-level contributions to precision electroweak observables. At one-loop, oblique corrections to the electroweak gauge boson propagators induced by diagrams involving the $T$ are given by~\cite{PEW}
\beqa
S_T &=& \frac{\s^2}{2\pi}\,\left[ \left(\frac{1}{3}- \c^2\right)\,\log \r \,+\,\c^2 \frac{(1+\r)^2}{(1-\r)^2}\,+\,\frac{2\c^2\r^2(3-\r)\log \r}{(1-\r)^3}\,-\frac{8\c^2}{3}
\right]\,, \CR
T_T &=& \frac{3}{16\pi}\,\frac{\s^2}{s_w^2c_w^2}\,\frac{m_t^2}{m_Z^2}\,\left[\frac{\s^2}{\r}-1-\c^2-\frac{2\c^2}{1-\r}\log \r \right]\,,\CR
U_T &=& -\frac{\s^2}{2\pi}\,\left[ \s^2\,\log \r \,+\,\c^2\frac{(1+\r)^2}{(1-\r)^2}\,+\,\frac{2\c^2\r^2(3-\r)\log \r}{(1-\r)^3}\,-\frac{8\c^2}{3}
\right] \,,
\eeqa{exactSTU}
where $\r=m_t^2/\mT^2$, and $s_w$ is the sine of the Weinberg angle. In addition, there is a contribution due to the shift of the Higgs couplings to the electroweak gauge bosons from their SM values~\cite{NSH}:
\beqa
S_h &=& - \frac{1}{3\pi} \frac{m_W^2}{g^2f^2}\,\log \frac{m_h}{\Lambda}\,,\CR
T_h &=& \frac{3}{4\pi c_w^2} \frac{m_W^2}{g^2f^2}\,\log \frac{m_h}{\Lambda}\,,
\eeqa{HiggsST} 
where $\Lambda$ is the scale where the Higgs loops are cut off. We will assume $\Lambda=4\pi f$. Furthermore, the operators induced by the new physics at scale $\Lambda$, given in Eq.~\leqn{d6ops}, contribute~\cite{Witek}
\beqa
S_{\rm UV} &=& \frac{4 c_s m_W^2}{\pi g^2 f^2}\,,\CR
T_{\rm UV} &=& - \frac{c_t m_W^2}{2 \pi e^2 g^2 f^2}\,.
\eeqa{uvST} 
The only important non-flavor-universal correction is the top-partner loop contribution to the $Zb_L\bar{b}_L$ vertex. To leading order in the limit $m_T\gg m_t\gg m_W$, this is given by~\cite{PEW}
\beq
\delta g_L^{b\bar{b}}=\frac{g}{c_w}\frac{\alpha}{8\pi s_w^2} \frac{m_t^4}{m_W^2\mT^2}
\frac{\lambda_1^2}{\lambda_2^2} \log \frac{\mT^2}{m_t^2}\,.
\eeq{zbb}
The correction to the $Zb_R\bar{b}_R$ vertex is negligible since it is not enhanced by the top Yukawa coupling.

\begin{figure}[tb]
\begin{center}
\centerline {
\includegraphics[width=2.2in]{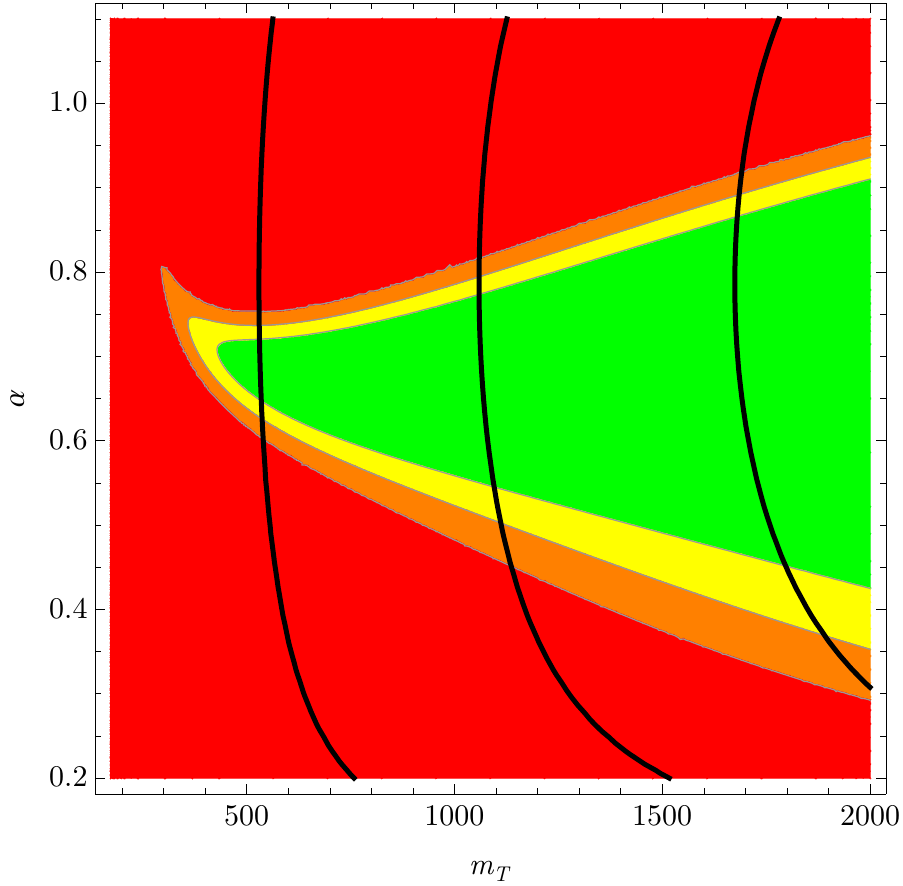}
\includegraphics[width=2.2in]{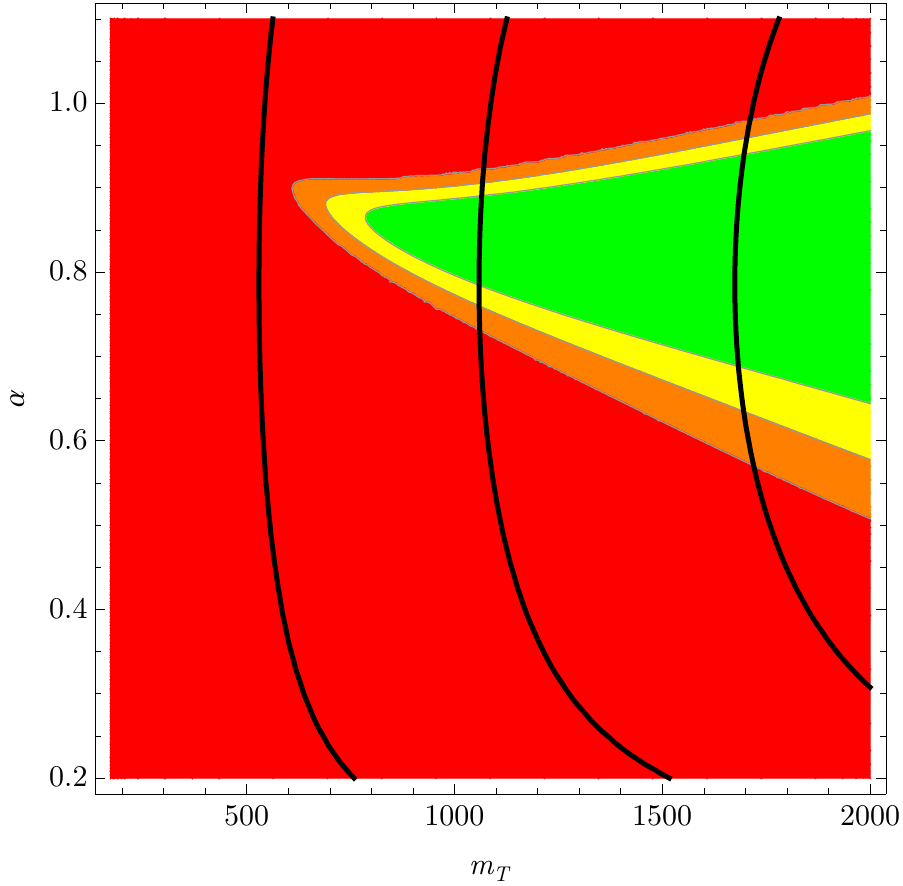}
\includegraphics[width=2.2in]{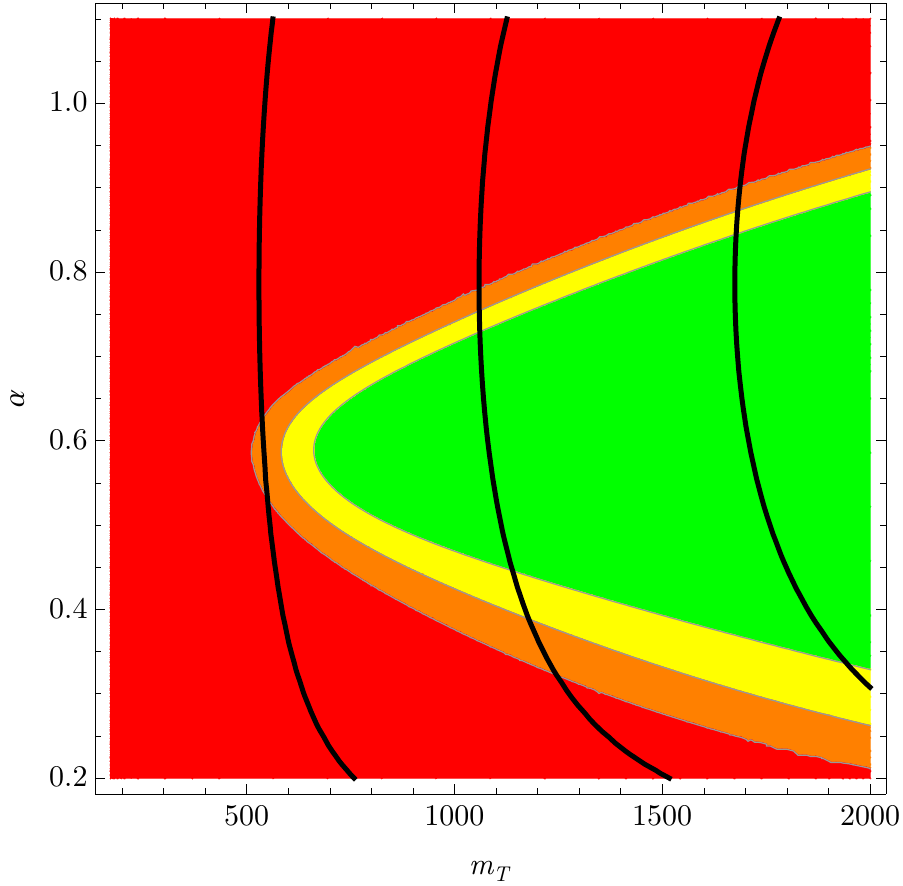}
}
\vspace{.2in}
\caption{Precision electroweak constraints on the minimal fermionic top partner model, in the $(m_T, \alpha)$ plane. The three panels display the variation of the bounds depending on the coefficients of the UV operators: (a) $c_s=c_t=0$; (b) $c_s=c_t=1$; (c) $c_s=+1$, $c_t=-1$. Thick black lines represent constant fine-tuning contours: from left to right, 20\%, 5\%, and 2\% fine tuning.}
\label{fig:PEW}
\end{center}
\end{figure}

The results of a fit to the precision electroweak observables~\cite{PDG} are shown in Fig.~\ref{fig:PEW}, where we also included contours of constant fine-tuning computed according to Eq.~\leqn{FT}. We conclude that:

\begin{itemize}

\item The lower bound on the top partner mass from precision electroweak observables is approximately 500 GeV;

\item The corresponding minimum level of fine-tuning on the Higgs mass is about 20\%. This is significantly better than in the MSSM with a 125 GeV Higgs, and comparable to the NMSSM with large $\lambda$~\cite{SUSY_Higgs_FT};

\item These conclusions do not depend strongly on the operators induced by the UV completion of the model, as long as the size of these operators is roughly consistent with naive dimensional analysis.

\end{itemize}

\subsection{Flavor Constraints}

By selecting the top quark to be the only one with a partner, and introducing mixing between the SM top and its partner, our model explicitly breaks the approximate flavor symmetry of the SM, leading to potential constraints from flavor-changing processes. We investigate these constraints in this section.

Including the mixing between the three SM generations, the mass terms of the up-type quarks in the gauge basis form a  $4\times4$ matrix $M_u^{IJ}$, while the down-type mass terms are described by a $3\times3$ matrix $M_d^{ij}$. (Here and below, capital indices run from $1$ to $4$, and the lower case indices from $1$ to $3$.) Diagonalizing these matrices requires 
\beqa
M_u \rightarrow L_u M_u R_u^\dagger\,,\CR
M_d \rightarrow L_d M_d R_d^\dagger\,,
\eeqa{rotation}
where $L$ and $R$ matrices rotate the left-handed and right-handed quark fields, respectively. The charged-current interactions in the gauge basis have the form
\beqa
{\cal L}_{c.c.} &=& g W^-_\mu J^{+\mu} + {~\rm c.c.}\,,\CR
J^{+\mu} &=& \frac{1}{\sqrt{2}} \bar{U}_L^I \gamma^\mu (P_3)_I^j (D_L)_j\,,
\eeqa{cc}
where 
\beq
(P_3)_I^j \equiv \left( \begin{array}{c} 
{1}_{3 \times 3} \\ \vec{0} \end{array} \right).
\eeq{P3}
In the mass basis, the charged current becomes
\begin{equation}
J^+_\mu = \frac{1}{\sqrt{2}} \bar{u}_L^I \gamma^\mu (L^\dagger_u)_I^J (P_3)_J^k (L_d)_k^l (d_L)_l\,,
\end{equation}
so that the generalization of the CKM matrix in our model is
\beq
(\tilde{V}_{CKM})_I^j =  (L^\dagger_u)_I^K (P_3)_K^k (L_d)_k^j.
\eeq{Vckm}
The elements of this matrix should in principle be determined by a fit to data. We will not attempt such a fit here. Since the SM CKM matrix provides an excellent description of flavor-changing processes for the first two generations and the $b$ quark, we assume the following structure:
\beq
\tilde{V}_{CKM} = \left( \begin{array}{ccc}
V_{ud} & V_{us} & V_{ub} \\
V_{cd} & V_{cs} & V_{cb} \\
c_\beta V_{td} & c_\beta V_{ts} & c_\beta V_{tb} \\
-s_\beta V_{td} & -s_\beta V_{ts} & -s_\beta V_{tb}
\end{array} \right)
\eeq{Vckm1}
where $V_{ij}$ are SM CKM elements. With this assumption, all flavor-violating new physics effects in $K$ and $B$ systems appear at loop-level only.   

Unlike the SM, rotations~\leqn{rotation} induce tree-level flavor-changing neutral currents (FCNC) in the left-handed sector~\cite{Lee:2004me}, since the weak-singlet $U_L$ mixes with the SM up-type quarks. The $Z$ boson couples to the current
\beq
J^\text{NC}_\mu = (\bar{U}_L)^I \gamma^\mu \left( T_3 - s_w^2 Q \right)_I^J (U_L)_J,
\eeq{Jnc1}
where 
\beq
(T_3)_I^J = \left( \begin{array}{cc} 
{1}_{3 \times 3} & 0 \\ 0 & 0\end{array} \right).
\eeq{T3}
Rotation to the mass basis yields flavor-changing couplings, proportional to
\beq
V_{\rm FCNC} = L^\dagger_u \left( T_3 - s_w^2 Q \right) L_u\,.
\eeq{treeFCNC}
These can generate tree-level contributions to rare $D$ meson decays and anomalous $D_0 - \bar{D}_0$ mixing, and flavor-changing top decays. Such contributions are however completely absent if 
\beq
L_u = \left( \begin{array}{ccc} 
{1}_{2 \times 2} & \ & \ \\ \ & c_\beta & -s_\beta \\ \ & s_\beta & c_\beta \end{array} \right)\,,
\eeq{Uumat}
since the only flavor-violating $Z$ coupling in this case is $ZtT$. 
Eq.~\leqn{Vckm} then requires $L_d = V_{CKM}^{SM}$. We will assume this texture in our analysis. Note, however, that due to large theoretical uncertainties associated with the $D$ system and the highly suppressed rates for anomalous top decays, significant deviations from this texture can still be consistent with experimental constraints~\cite{Lee:2004me}.  

At the one-loop level, our model predicts new contributions to $\Delta F=2$ and $\Delta F=1$ processes in $B$ and $K$ systems. Let us first consider $\Delta F=2$. The effective Hamiltonian that governs the $B^0_s-\bar{B}^0_s$ system is
\beq
{\mathcal H}_{B_s} = \frac{G_F^2}{16 \pi^2} M_W^2 \sum_{I,J=u,c,t,T} \lambda_I \lambda_J F ( x_I, x_J ; M_W ) \times (\bar{b} s)_{(V-A)} (\bar{s} b)_{(V-A)},
\eeq{Heff}
where we defined $\lambda_I \equiv V^*_{Ib} V_{Is}$ and $x_I \equiv M_I^2/M_W^2$. The $F$ functions are given in the Appendix B. Hamiltonians for the $K^0$ and $B^0_d$ systems are obtained by simple substitutions, $b\to d$ and $s\to d$, respectively. At leading order in $v/f$ expansion, our results agree with Refs.~\cite{F2Jay,F2Buras,BurasAll}; however, our expressions are exact in $v/f$. To a good approximation, the size of the new physics effects in $\Delta F=2$ observables can be estimated as the fractional deviation of the Wilson coefficient in Eq.~\leqn{Heff} from its SM value. (This estimate does not take into account some effects, such as the running of the Wilson coefficient between the scales $m_T$ and $m_t$, which are however expected to be small.) We find that the maximum deviations on the parameter space of our model are: 0.5\% for $\Delta m_K$; about 20\% for $\epsilon_K$; and about 35\% for $|\Delta m(B_d)|$ and $|\Delta m(B_s)|$. Such deviations are currently easily allowed by data: see, for example,~\cite{UTfit}. 

\begin{figure}[tb]
\begin{center}
\centerline {
\includegraphics[width=2.5in]{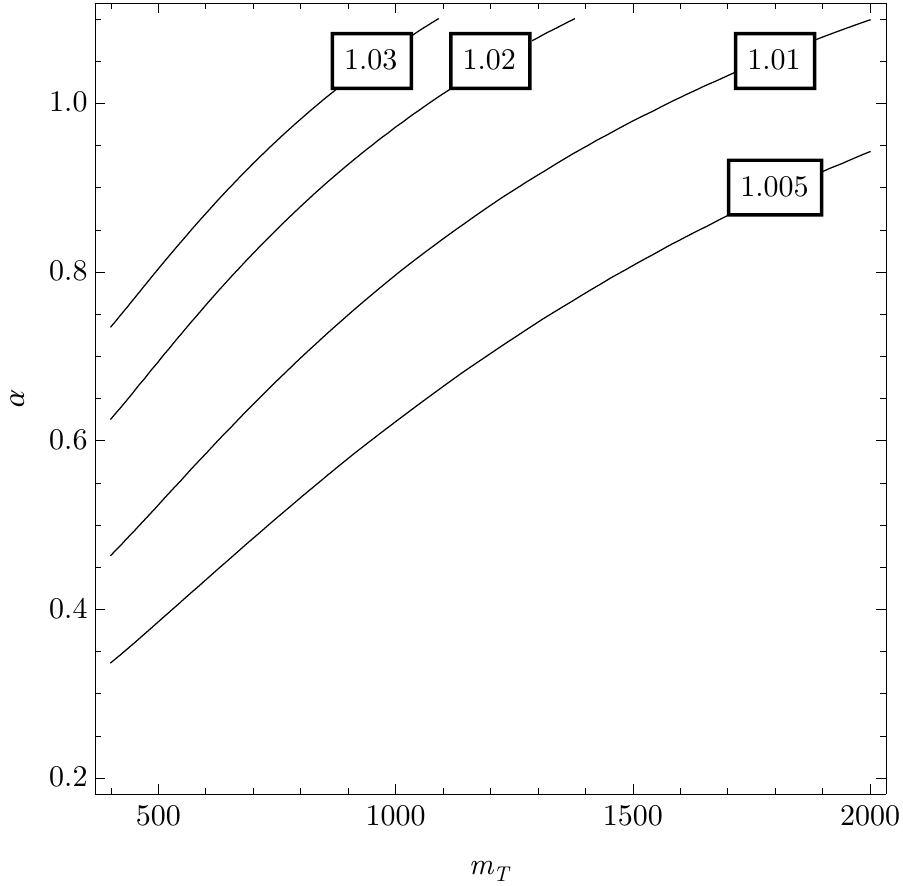}
\includegraphics[width=2.5in]{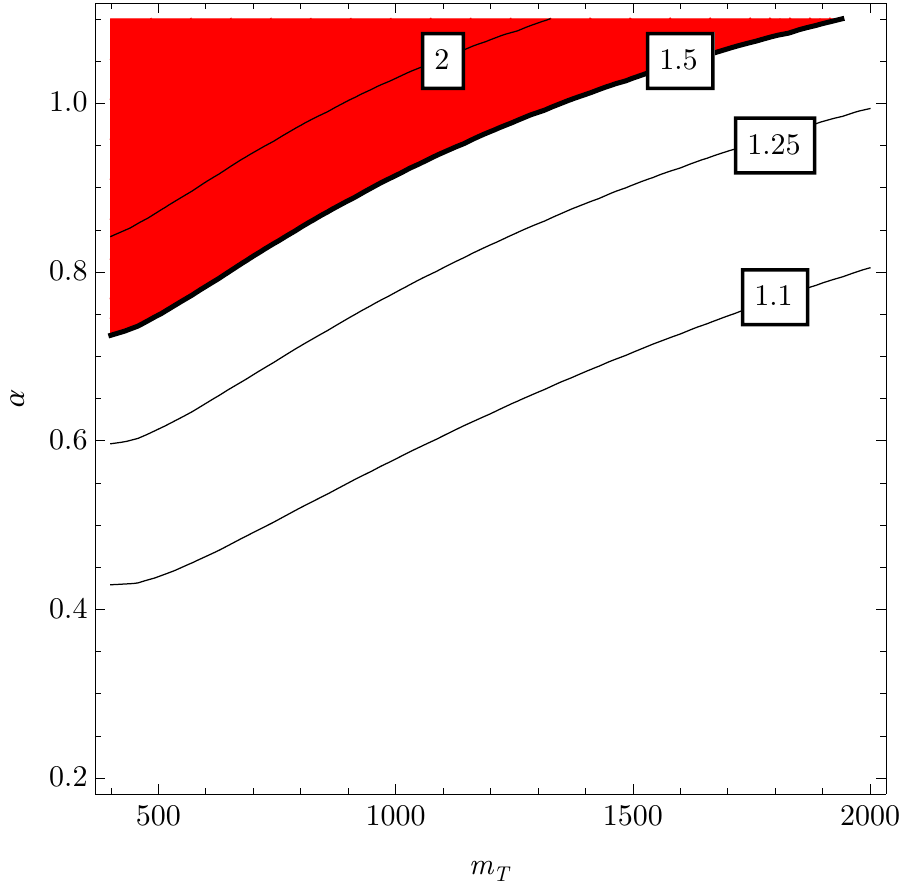}
}
\vspace{.2in}
\caption{Fractional deviations of the $\bar{B}\to X_s\gamma$ (left panel) and $B_s\to\mu^+\mu^-$ (right panel) branching ratios from the SM predictions. The thick line on the right panel corresponds to the LHCb upper bound on Br$(B_s\to\mu^+\mu^-)$; all points above the line are now ruled out.}
\label{fig:raredecays}
\end{center}
\end{figure}

We next consider the two most constrained $\Delta F = 1$ decays, $b\to s\gamma$ and $B_s\to\mu^+\mu^-$. The $b\to s\gamma$ amplitude is proportional, in the leading-log approximation, to the coefficient $C_7$ of the operator $P_7=\frac{e}{16\pi^2}m_b(\tilde{s}_L \sigma^{\mu\nu}b_R) F_{\mu\nu}$, evaluated at the scale $m_b$. The top-quark contribution to this coefficient is given by
\beq
X_t = -\frac{1}{2} A_0^t(x_t) \eta^{16/23} - \frac{4}{3}F_0^t(x_t)\left( \eta^{14/23}-\eta^{16/23}\right),
\eeq{SMbsg}
where the functions $A_0^t(x)$ and $F_0^t(x)$ can be found in Appendix B, and $\eta=\alpha_s(m_t)/\alpha_s(m_b)$. The only effect of the top partner is to replace
\beqa
A_0^t(x_t) &\to& c_\beta^2 A_0^t(x_t) + s_\beta^2 A_0^t(x_T), \CR
F_0^t(x_t) &\to& c_\beta^2 F_0^t(x_t) + s_\beta^2 F_0^t(x_T), 
\eeqa{AFswap}
in these expressions. (The first term in the $v/f$ expansion of these formulas agrees with Refs.~\cite{BSgamma,BurasAll}; however, our formulas are exact in $v/f$.) The resulting deviations of the $b\to s\gamma$ branching ratio from the SM are shown in the left panel of Fig.~\ref{fig:raredecays}. In the region of interest, the deviations are at most about 5\%. Given that both the experimental measurement~\cite{PDG} and the NNLO SM theoretical prediction~\cite{SMbsg} have uncertainties between 5 and 10\%, such deviations cannot be currently ruled out. The right panel of the figure shows the deviation of the $B_s\to\mu^+\mu^-$ branching ratio from the SM prediction, evaluated using the formulas given in Ref.~\cite{Bmumu}. We also indicate the region ruled out by the recent LHCb bound~\cite{LHCb}, Br$(B_s\to\mu^+\mu^-)<4.5\times 10^{-9}$ at 95\% c.l., which is only a factor of 1.5 above the SM prediction. This is the strongest current bound on the top partner from flavor physics, even though it is still weaker than precision electroweak constraints. Notice that the results of Ref.~\cite{Bmumu} are valid to leading order in the $v/f$ expansion. Given the potential importance of this bound, a more precise calculation is desirable.

\subsection{Direct Searches at the LHC}

\begin{figure}[tb]
\begin{center}
\centerline {
\includegraphics{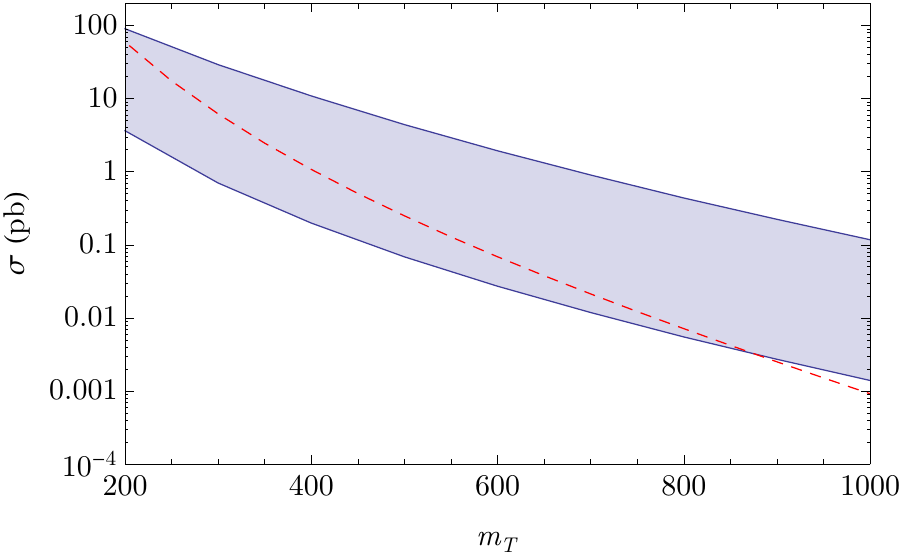}
}
\vspace{.2in}
\caption{Production cross section of top partners at the 7 TeV LHC. For pair-production (red), we use {\tt Hathor v1.2}~\cite{TT_nnlo} to calculate at NNLO in QCD.  For single production (blue), we use {\tt MadGraph5 v1.3.32}~\cite{TQ_lo} to calculate at LO.  The single production cross-section depends on $\alpha$, with the band indicating the cross-sections for $0.2 < \alpha < 1.1$.}
\label{fig:xsec}
\end{center}
\end{figure}

The two production mechanisms for the top partner are strong pair-production, $q\bar{q}/gg \to T\bar{T}$, and electroweak single production, $b q \to T q^\prime$ or $q q^\prime \to T b$. The production cross sections at the 7 TeV LHC are shown in Fig.~\ref{fig:xsec}. For pair-production the cross-section is calculated at NNLO in QCD using {\tt Hathor v1.2}~\cite{TT_nnlo}, with renormalization and factorization scales set to the top partner mass. For single-production the cross-section is calculated at LO using {\tt MadGraph5 v1.3.32}~\cite{TQ_lo}. While an NLO calculation of single-top production cross section is  available~\cite{NLO_1t}, its use is not justified in out study due to large model uncertainty in the $bTW$ coupling. In this case, we use the {\tt MadGraph} default setting for renormalization and factorization scale, variable event-by-event. (For both pair- and single-production, varying renormalization and factorization scales within a factor of 2 leads to at most a few $\%$ variations in the cross sections.) At the 7 TeV LHC, due to the relatively small phase space for producing heavy particles, single production overcomes its electroweak suppression and can be comparable to pair production.

Decay channels of the top partner include $th$, $tZ$ and $bW$~\cite{Han,PPP}. In the limit $f\gg v$, the branching ratios are 25\%, 25\%, and 50\%, respectively, as can be easily seen from the Goldstone boson equivalence theorem. An explicit calculation of the partial widths yields~\cite{PPP}:
\beqa  
& & \hskip-0.5cm \Gamma(bW) = \frac{g^2 \s^2 m_T^3}{64\pi m_W^2} \,f(x_W, x_b)\,g(x_b, x_W),\CR
& & \hskip-0.5cm \Gamma(tZ) = \frac{e^2 \s^2 \c^2 m_T^3}{128\pi c_w^2 s_w^2 m_Z^2} \,f(x_Z, x_b)\,g(x_b, x_Z),\CR
& & \hskip-0.5cm \Gamma(th) = \frac{m_T}{64\pi} \, f(x_t, x_h) \left[ ( 1+x_t^2-x_h^2) \left( (C^L_{Tth})^2 + (C^R_{Tth})^2 \right)+ 4 C^L_{Tth} C^R_{Tth} x_t \right],
\eeqa{widths}
where $x_i\equiv m_i/m_T$, the kinematic functions are defined as
\beqa
f(x_i,x_j) &=& \sqrt{(1-(x_i+x_j)^2)(1-(x_i-x_j)^2)}\,,\CR
g(x_i,x_j) &=& 1-x_i^2+x_j^2(1+x_i^2) - 2 x_j^4\,,
\eeqa{fandg}
and the constants appearing in the $tTh$ vertex are given in Appendix A.

There exist several searches for vector-like top partners at CMS and ATLAS~\cite{CMS_1,Chatrchyan:2011ay,CMS_3,CMS_4,Aad:2012bt,Aad:2012xc}. These searches focus on pair production and on one particular decay mode of the top partner, either $bW$ or $tZ$, and assume $100\%$ branching fraction to that mode. In our model, the signal is generally a mixture of pair and single production, and multiple decay channels are possible. As a result, the bounds on the top partner masses obtained by CMS and ATLAS are not directly applicable, but it is possible to ``recast" the published analyses to estimate the bounds in our model.~\footnote{See~\cite{Rao:2012gf} for a recent theoretical analysis that recasts experimental searches in terms of limits on top-partners with general values of the branching fractions to $bW$, $tZ$, and $th$.}   Below we present such an estimate, based on the CMS search in the $bbWW$ final state with $5.0$ fb$^{-1}$ integrated luminosity~\cite{CMS_1}.  In the interesting parameter space of our model, the dominant decay mode for the 
$T$ is $bW$, making $bbWW$ searches most sensitive.  Furthermore, the CMS analysis places the strongest bounds as it is updated to use the full 2011 dataset.

\begin{table}[tb]
  \centering
  \begin{tabular}{c c c c}
    \hline
    Final state & Raw $\text{Br}_\ell \epsilon_i A_i$ (\%) & Rescaled $\text{Br}_\ell \epsilon_i A_i$ (\%) & $\text{Br}_i \text{Br}_\ell \epsilon_i A_i$ (\%) \\
    \hline
    $bWbW$ & 0.36& 0.29 & 0.12 \\
    $bWtZ$ & 0.034 & 0.027 & 0.0046 \\
    $bWtH$ & 0.022 & 0.018 & 0.0011 \\
    $tZtZ$ & 0.0015 & 0.0012 & $8.5 \times 10^{-5}$ \\
    $tHtH$ & $9.9 \times 10^{-4}$ & $7.9\times10^{-4}$ & $7.5 \times 10^{-6}$ \\
    \hline
  \end{tabular}
  \caption{Estimated raw $\text{Br}_\ell \epsilon_i A_i$, rescaled $\text{Br}_\ell \epsilon_i A_i$ and $\text{Br}_\ell \text{Br}_i \epsilon_i A_i$ for the various decays of pair produced $T\overline{T}$.  All values assume $m_T = 400~{\rm GeV}$ and $\alpha = \pi/4$.  See text for the definition of raw and rescaled efficiencies.  Here $\text{Br}_\ell$ denotes the dileptonic branching fraction for $WW$, which is common to all decay modes.}
  \label{tab:acceffbr}
\end{table}

The number of signal events expected in a given analysis can be written as
\beq
N_{\rm sig} = \sum_{ij} \sigma_i \mathcal{L} {\rm Br}_{ij} \epsilon_{ij} A_{ij},
\eeq{bfeff}
where $\sigma_i$ is the cross-section for each production channel, $\mathcal{L}$ is the integrated luminosity used in the search, ${\rm Br}_{ij}$ is the branching fraction for an event produced via channel $i$ to result in the final state $j$ after the decay of all unstable particles, $\epsilon_{ij}$ is the efficiency for detecting the final state $j$ in a given analysis, and $A_{ij}$ is the acceptance for the final state $j$. (Note that $\epsilon$ and $A$ depend on the production channel, since final-state particles have different kinematic distributions depending on the production mechanism.) The efficiency and acceptance for the particular production and decay mode assumed in the CMS analysis ($T\bar{T}$, with $T\to bW$) can be found in~\cite{CMS_1}. We estimated the $\epsilon A$ of all other relevant final states by modeling the acceptance and selection cuts of~\cite{CMS_1} on a sample of Monte Carlo (MC)-generated $T\bar{T}$ events. (Since the CMS analysis required two isolated leptons, and vetoed events with dilepton invariant mass close to the $Z$ boson, the efficiencies for events with a single $T$ to pass the cuts are extremely small, and we did not include the single production channel in our analysis.) For this estimate, we generated parton-level events using {\tt MadGraph 5 v1.3.32}~\cite{TQ_lo}, showered and hadronized them using {\tt Pythia 6.426}~\cite{pyth}, and applied simplified detector simulation using {\tt PGS 4.0}~\cite{pgs}. Unfortunately, {\tt PGS 4.0} significantly underestimates the efficiency of b-tagging, compared to the TCHEM algorithm used by the CMS in this analysis.\footnote{This can be easily seen by comparing PGS and TCHEM efficiencies on an SM $b\bar{b}$ sample. The peak efficiencies are 0.4 for PGS and 0.7 for TCHEM.} To address this issue, we ignored the b-tag information provided by {\tt PGS}, and instead applied $p_T$-dependent TCHEM efficiencies~\cite{TCHEM} to the b-jets in our sample. This procedure yields the ``raw" $\epsilon A$ values for all possible final states, as a function of $m_T$ and $\alpha$. For example, values of  $\epsilon A$ for $m_T=400$ GeV, $\alpha=\pi/4$ are listed in the first column of Table~\ref{tab:acceffbr}.
The MC simulation and analysis procedure was validated on a sample of events with the final state considered by CMS, $WbWb$ with 2 leptonic $W$'s.  We found that the $\epsilon A$ determined using our simulation for a $400~{\rm GeV}$ top partner is $0.36\%$, compared with the value of $0.29\%$ quoted in the CMS analysis. Given the crude nature of our MC simulations, this level of agreement is very reasonable. Even so, in deriving the bounds, we rescale the raw MC estimates of ${\rm Br} \epsilon A$ by a correction factor of $0.29/0.36$; in other words, we use the $\epsilon A$ quoted by CMS for the $WbWb$ channel, and use the MC to estimate the relative $\epsilon A$ of other channels with respect to  $WbWb$. The resulting estimates are collected in Table~\ref{tab:acceffbr}. It is clear that the rates of events from final states other than $WbWb$ that pass the analysis cuts are quite small. While our estimates of those rates suffer from significant systematic uncertainties due to the crude detector simulation used, it is reassuring that even if the rates were inflated by a factor of two they would remain subdominant. Thus, our bounds on the top partner mass are robust. 

\begin{figure}[tb]
  \centering
  \includegraphics{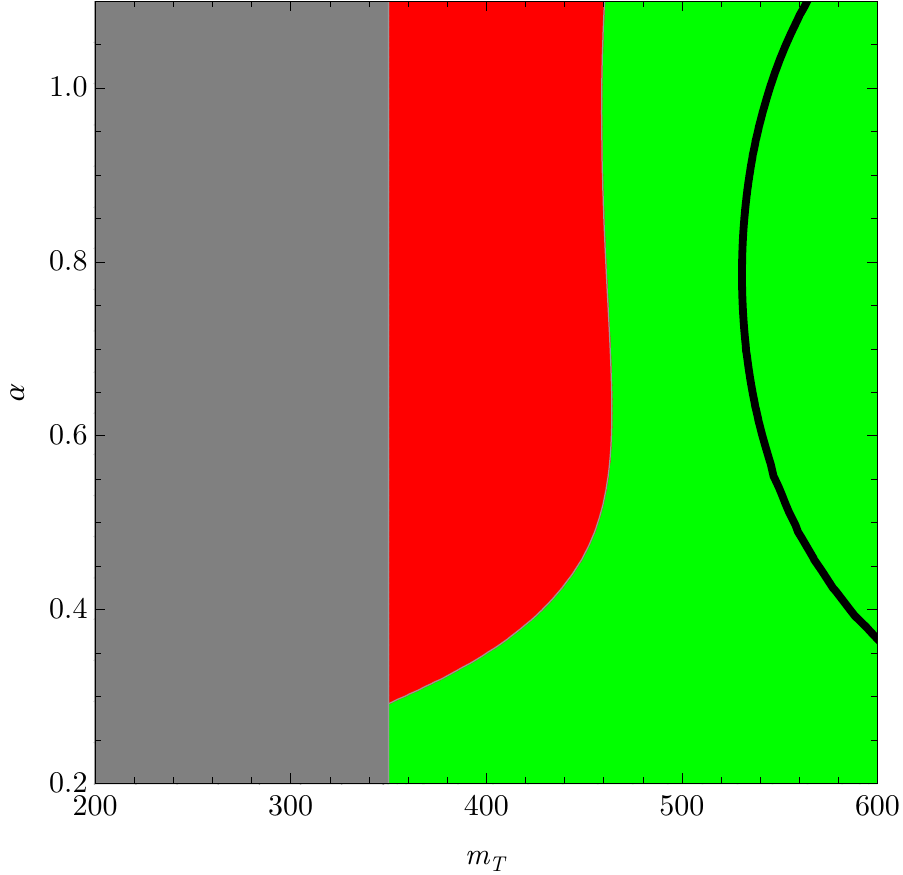}
  \caption{The estimated 95\% excluded region from the CMS $T\to Wb$ search~\cite{CMS_1} in terms of $M_T$ and $\alpha$ is shown as the red contour.  The green is the allowed region.  The thick black line corresponds to 20\% fine-tuning.  The CMS analysis does not quote efficiencies below top partner masses of $350$ GeV, so no bound is shown (grey region).   This low-$m_T$ region is in any case already ruled out by precision electroweak fits.}
  \label{fig:direct-excl}
\end{figure}

The estimated $95\%$ C.L.\ exclusion region as a function of $m_T$ and $\alpha$ is presented in Fig.~\ref{fig:direct-excl}.
The typical bound on the top partner mass is about 450 GeV, with somewhat weaker bounds for small $\alpha$.\footnote{A reanalysis of the published LHC searches in the context of the ``Bestest" Little Higgs model appeared recently in Ref.~\cite{BestestLHC}, where similar bounds on the top partner mass were found.}  
It is clear that direct collider searches are just beginning to probe the region that is not already ruled out by precision electroweak constraints. Note that the least fine-tuned parameter space regions will be probed by direct LHC searches in 2012. 

ATLAS has also published a search for a singly-produced vector-like heavy quark, in particular using the channel $pp \to Qq \to Wqq^\prime$~\cite{ATLAS_single}. The lower bound on the $Q$ mass of order 900 GeV was reported. This search potentially has sensitivity to our model, since single top partner production contributes to this final state. However, the $Q$ in the ATLAS analysis was assumed to have direct coupling to first-generation quarks, while in our model $T$ only couples to the third generation, resulting in much lower production cross sections. As a result, this analysis does not yet put interesting bounds on the top partner mass. For example, ATLAS sets a bound on $\sigma(pp \to Qq)\times$Br$(Q\to Wq)$ of about 2 pb for a 500 GeV $Q$ mass; in our model, a single top-partner production cross section of this size only occurs at the upper edge of the band shown in Fig.~\ref{fig:xsec}. This requires values of $\alpha\approx 1$, which are already ruled out by precision electroweak constraints. Note, however, that the published ATLAS analysis only uses 1.04 fb$^{-1}$ of data; updated versions of this analysis will have sensitivity to interesting regions of the top partner parameter space.

\section{Higgs Properties}
\label{sec:Higgs}

If the LHC evidence for the Higgs boson at 125 GeV is correct, detailed measurements of the Higgs production cross section and branching ratios should be possible within the next few years. In our model, these properties deviate from the SM predictions. There are two important effects. First, the $hWW$ and $hZZ$ couplings are shifted,\footnote{These shifts are due simply to the composite nature of the Higgs, and not to the presence of the top partners. They can be described within the framework developed in Refs.~\cite{CompH}. } leading to deviations in the $WW$ and $ZZ$ branching fractions and, via the $W$-loop contribution, in the Br($h\to\gamma\gamma$). Second,     
loops of top partners produce corrections to the $hgg$ and $h\gamma\gamma$ vertices, leading to deviations in the expected production cross section and, again, Br($h\to\gamma\gamma$). 

The production rate of $h$ via gluon fusion is proportional to $\Gamma(h\to gg)$. Assuming that gluon fusion is the dominant Higgs production mechanism, the rates $\sigma(pp\to h\to VV)$ in our model, normalized to their SM values, are
\beq
R_V = \frac{\Gamma(h\to gg) {\rm Br}(h\to VV)}{\Gamma_{\rm SM}(h\to gg) {\rm Br}_{\rm SM}(h\to VV)}\,,
\eeq{rate_corr1}
where $V=\gamma, Z, W$. The total Higgs decay rate at $m_h=125$ GeV is dominated by the $b\bar{b}$ mode. The bottom Yukawa coupling can be incorporated in our model as an explicit breaking of the global symmetry; this would not spoil naturalness due to the small numerical value of $y_b$. At leading order in $v/f$, this results in the $h\bar{b}b$ coupling identical to the SM value. There may be corrections at higher orders in $v/f$; however, their form is not fixed by the symmetry, and is model-dependent. If they are ignored, we simply get
\beq
R_V = \frac{\Gamma(h\to gg) \Gamma(h\to VV)}{\Gamma_{\rm SM}(h\to gg) \Gamma_{\rm SM}(h\to VV)}\,.
\eeq{rate_cor2r}
Note that the dropped terms in the $h\bar{b}b$ vertex are potentially of the same order as the corrections to the $hgg$ and $hVV$ couplings, so these predictions have an inherent ${\cal O}(1)$ ambiguity. Still, we compute them as an indication of the likely size of the effect. We should also note that our predictions for {\it ratios} of rates, such as for example $R_\gamma/R_W$, are free of this ambiguity.

The ratios of the $h\to WW$ and $h\to ZZ$ decay rates to the SM predictions are given by
\beq
\frac{\Gamma(h\to W^+W^-)}{\Gamma_{\rm SM}(h\to W^+W^-)}\,=\,\frac{\Gamma(h\to ZZ)}{\Gamma_{\rm SM}(h\to ZZ)}\,=\,
1-\frac{2m_W^2}{g^2f^2}\,.
\eeq{ww_shift}
The $h\to\gamma\gamma$ decay rate is~\cite{HLMW,LHDM}
\beq
\Gamma(h\to\gamma\gamma) = \frac{\alpha^2g^2}{1024\pi^3}\,\frac{m_h^3}{m_W^2}\,\left| F_1(\tau_W) \,+\,
\sum_{i\not= t} Q_i^2 N_{c,i} F_{1/2}(\tau_i) \,+\, 3 Q_t^2 {\cal A}_{\rm top} \right|^2\,,
\eeq{hgg}
where $\tau_i = 4m_i^2/m_h^2$; the sum runs over all SM fermions except the top; $Q_i$ is the electric charge of the $i$-th fermion and $N_{c,i}$ its color multiplicity (3 for quarks, 1 for leptons). The top contribution to the decay amplitude in our model is given by
\beq
 {\cal A}_{\rm top}  = \frac{\sqrt{2}m_W}{gm_t} \left( C_{tth} F_{1/2}(\tau_t) + \frac{m_t}{m_T} C_{TTh} 
F_{1/2} (\tau_T)\right)\,,
\eeq{hgg_tops}
where the constants $C$ are given in Appendix A; while in the SM, ${\cal A}_{\rm top}=F_{1/2}(\tau_t)$. Here we used the standard notation for the loop functions,
\beqa
F_1(x) &=& 2+3x+3x(2-x) f(x)\,,\CR
F_{1/2}(x) &=& -2x\left( 1+(1-x)f(x)\right) \,,\CR
f(x) &=& \left[ \sin^{-1}\left(\sqrt{\frac{1}{x}}\right)\right]^2~~{\rm if}~x > 1, \CR
& & 
-\frac{1}{4}\left[\log\left(\frac{1+\sqrt{1-x}}{1-\sqrt{1-x}}\right)- i\pi\right]^2~{\rm if}~x < 1.
\eeqa{Fs} 
The $h\to gg$ decay rate is given by~\cite{HLMW}
\beq
\Gamma(h\to gg) = \frac{\alpha_s^2g^2}{512\pi^3}\,\frac{m_h^3}{m_W^2}\,\left| 
\sum_{i\not= t} F_{1/2}(\tau_i) \,+\, {\cal A}_{\rm top} \right|^2\,.
\eeq{hglgl}

\begin{figure}[tb]
\begin{center}
\centerline {
\includegraphics[width=3in]{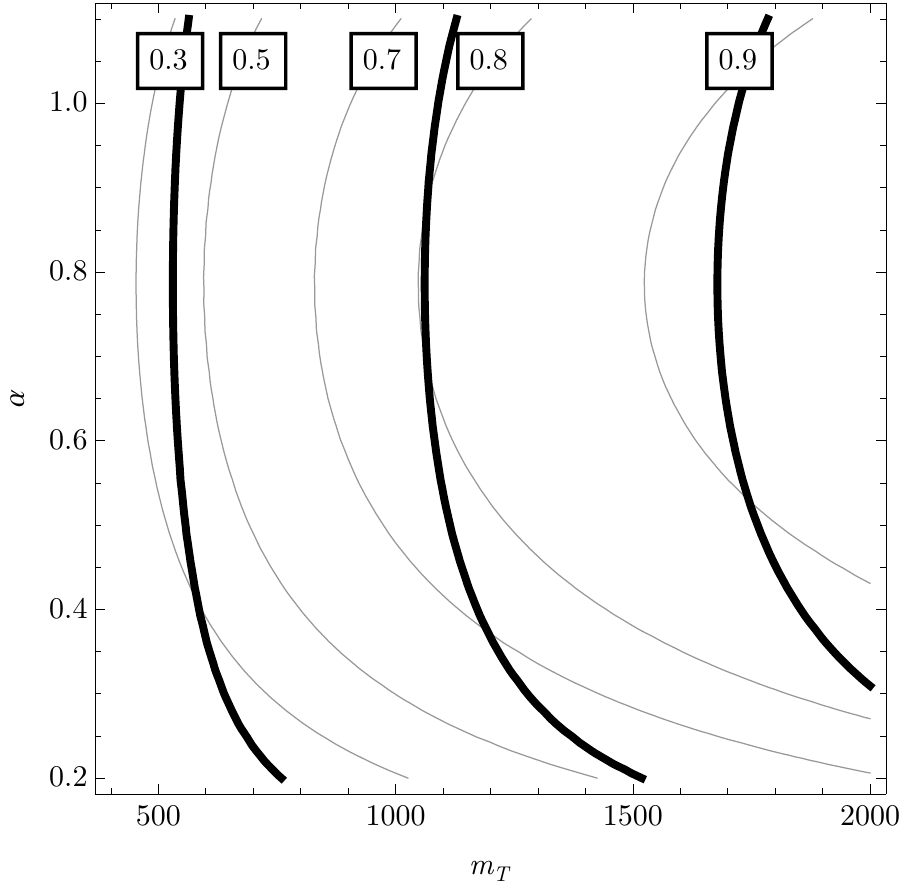}
\includegraphics[width=3in]{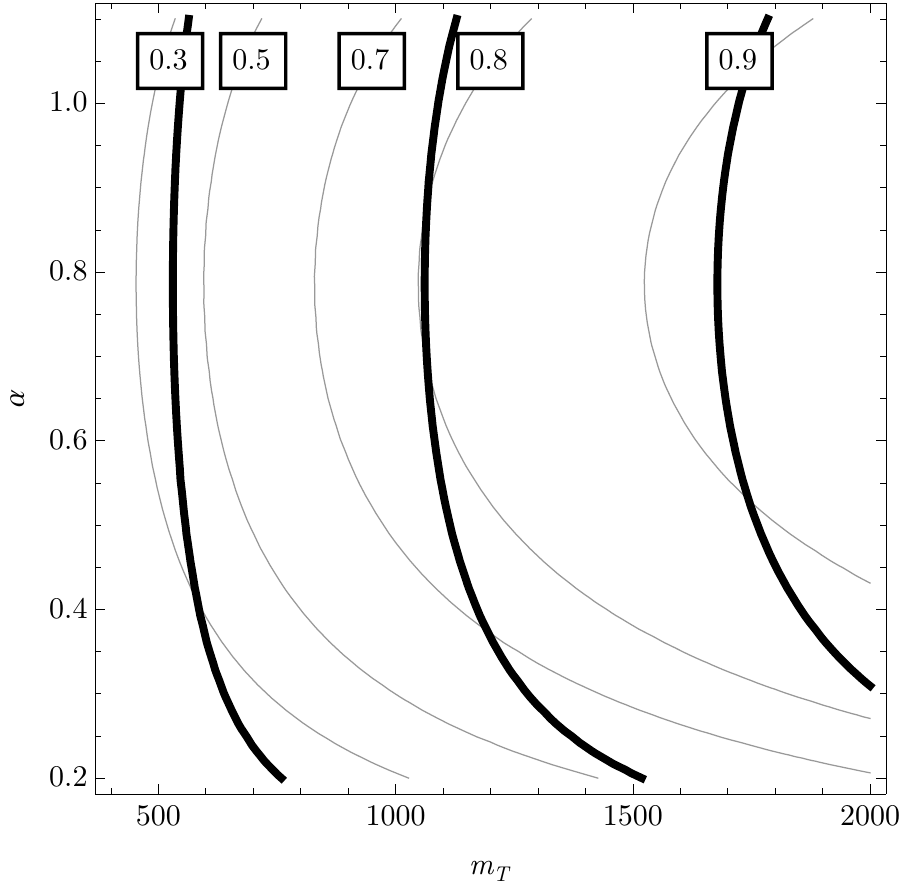}
}
\vspace{.2in}
\caption{Event rates for $h\to \gamma\gamma$ (left) and $h\to WW$ (right), normalized to the corresponding SM rates, for $m_h=125$ GeV. Thick black lines represent constant fine-tuning contours: from left to right, 20\%, 5\%, and 2\% fine tuning.}
\label{fig:Rs}
\end{center}
\end{figure}

The predicted rates $R_W=R_Z$ and $R_\gamma$ are shown in Fig.~\ref{fig:Rs}. Comparing with the precision electroweak constraints, we conclude that large suppression of the rates in both $WW/ZZ$ and $\gamma\gamma$ channels is possible: the rates can be as low as $30$\% of the SM prediction. Deviations are the strongest for the least fine-tuned regions of parameter space: for example, if we demand EWSB fine-tuning of 5\% or better, the minimal possible deviation in $R_W$ and $R_\gamma$ is 20\%. As noted above, these predictions should be taken with a grain of salt, since they can be modified by the model-dependent ${\cal O}(v/f)$ terms in the $hb\bar{b}$ coupling. Still, it is interesting that large, potentially observable deviations from the SM may occur throughout the natural parameter space. 

As remarked above, the ratio $R_\gamma/R_W$ provides a robust test of the structure since it's insensitive to the model-dependent embedding of the bottom Yukawa. Unfortunately, throughout the parameter space of our model, the deviations of this ratio from the SM prediction are well below 1\%, too small to be observed. The reason is that to a very good approximation, the fractional deviations of the $hWW$ coupling and the top loop contributions to $hgg$ and $h\gamma\gamma$ are the same. 

\if

There is another potential model-dependent contribution in both $R_\gamma$ and $R_W$, from new physics at scale $\Lambda$. This new dynamics is outside of the regime of validity of our model, and should be parametrized by effective operators suppressed by powers of $\Lambda$. The relevant operators are
\beq
\frac{\alpha_s}{\Lambda^2}(H^\dagger H) G_{\mu\nu}G^{\mu\nu},~~~\frac{\alpha}{\Lambda^2}(H^\dagger H) F_{\mu\nu}F^{\mu\nu}.
\eeq{Higgs_ops}
Operators of this size could be induced, for example, by loops of a new colored/electrically charged scalar, with mass of order $\Lambda$, if such scalar is strongly coupled to the Higgs. This is natural if global symmetry breaking is due to strong dynamics at scale $\Lambda$. After the Higgs acquires a vev, these two operators would contribute to $gg\to h$ and $h\to\gamma\gamma$ rates, respectively. These UV contributions are suppressed compared to the top partner effects computed above, by a factor of $\sim 4\pi$. We conclude that our predictions are robust in this sense.

\fi

\section{Top Properties}
\label{sec:top}

\begin{figure}[tb]
\begin{center}
\centerline {
\includegraphics[width=3in]{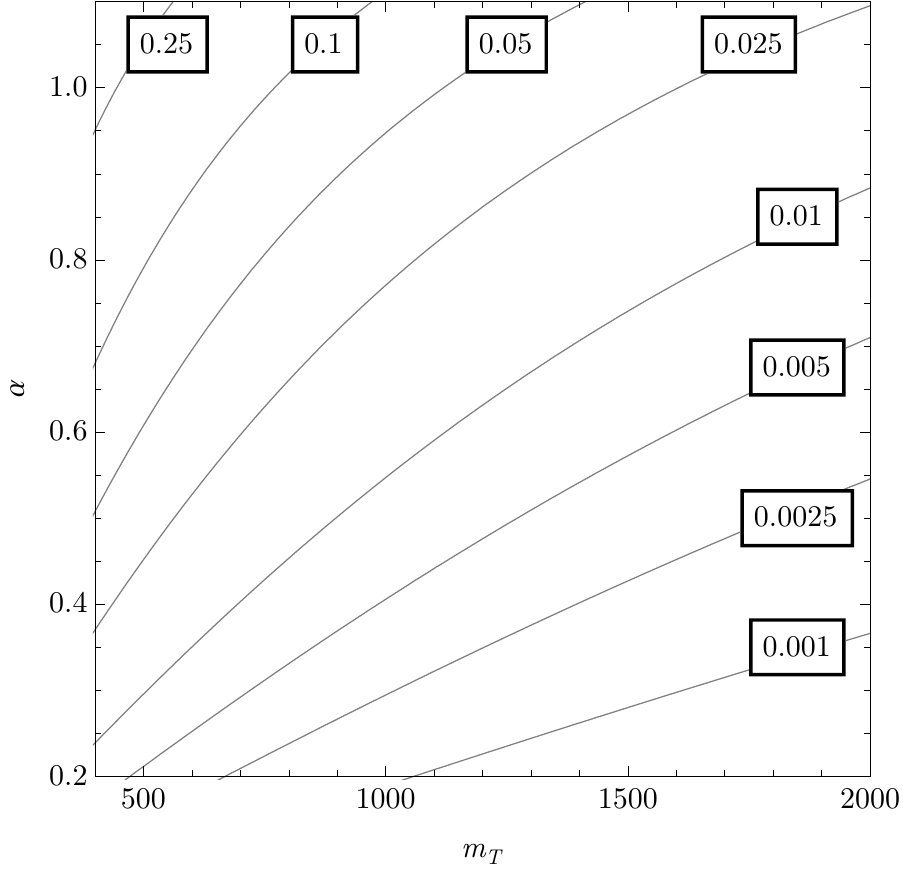}
\includegraphics[width=3in]{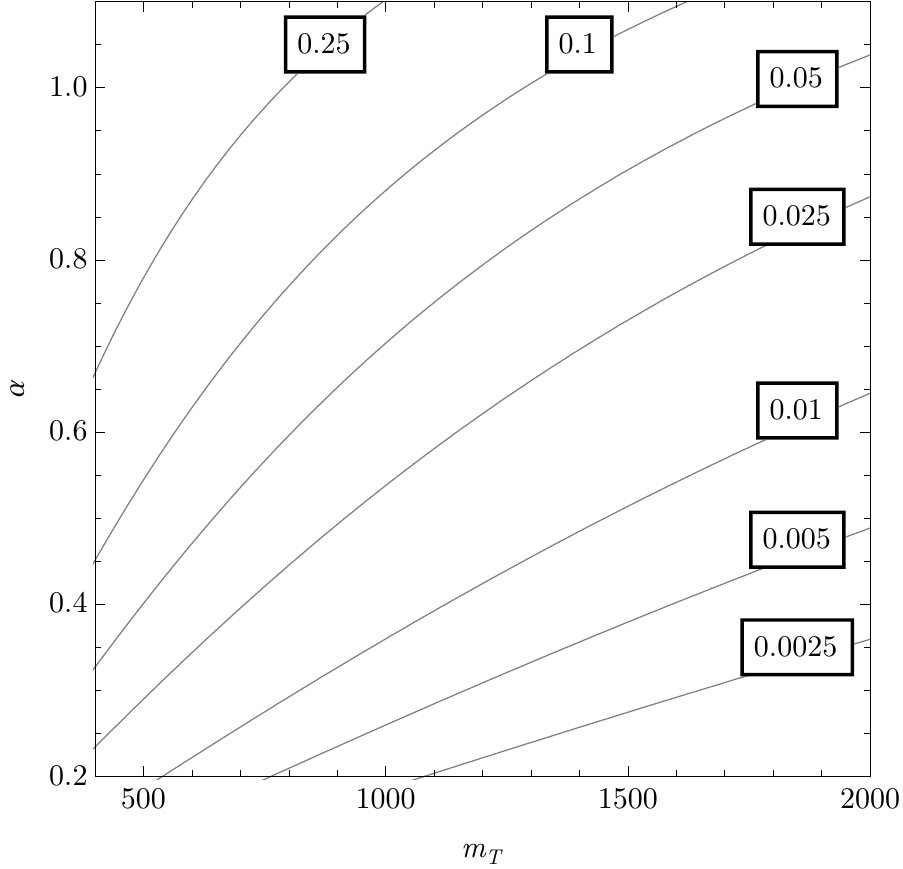}
}
\vspace{.2in}
\caption{Fractional deviations of the axial (left) and vector (right) components of the $t\bar{t}Z$ coupling from their SM values.}
\label{fig:topcoup}
\end{center}
\end{figure}

At order $v/f$, the lighter top eigenstate, which we identified with the SM top, actually contains an admixture of the $SU(2)$-singlet left-handed field $U_L$. As a result, the chiral structure of the top couplings to the $Z$ deviates from the SM predictions at this order. To quantify this effect, in Fig.~\ref{fig:topcoup} we plot the ratio of the vector and axial components of the $t\bar{t}Z$ coupling expected in our model, normalized to their SM values. Deviations of order 10\% or more in $g_A$, and up to 30\% in $g_V$, are possible in regions consistent with precision electroweak constraints. It was estimated that the 14 TeV LHC with 3000 fb$^{-1}$ integrated luminosity would be able to probe $g_A$ at the 5-10\% level and $g_V$ at the 15-30\% level~\cite{Baur}. A proposed 500 GeV linear electron-positron collider would reach precision of 2\% on $g_A$ and 5\% on $g_V$~\cite{ILC,BPP}. Though we would expect the top partner to be discovered in direct searches before these measurement become possible, they would still be of great interest to confirm the structure of the model.  

\section{Conclusions}
\label{sec:conc}

Naturalness, together with evidence that electroweak-symmetry breaking sector remains weakly coupled up to scales well above 1 TeV, implies that a light Higgs must be accompanied by new particles that cancel the quadratically divergent Higgs mass contribution from the SM top loop. In SUSY, these particles are scalar top (stop) quarks, and LHC phenomenology of stops has been a subject of much work recently. In this paper, we studied an alternative which has not received as much attention so far: naturalness restoration by spin-1/2 top partners. We focused on a minimal model where  
this mechanism is realized, which is essentially the top sector of the Littlest Higgs model. We explored current experimental constraints on this model from all relevant sources: precision electroweak fits, flavor physics, and direct LHC searches. We found that the current bound on the top partner mass is about 500 GeV, and is dominated by precision electroweak data, although direct searches are rapidly entering the hitherto allowed mass range. Given these bounds, accommodating a 125 GeV Higgs boson in this model requires only a modest level of fine-tuning, of order 20\%. Thus, we conclude that natural EWSB is possible in theories with sub-TeV-scale spin-1/2 top partners.

In the near future, direct searches for the top partners at the LHC will continue, gaining more sensitivity as more data is collected. The decay channels of the top partner include $bW$, $tZ$ and $th$, all of which have order-one branching ratios; this situation is not special to our model but is in fact quite generic. Also, while existing searches focus on pair-production of the top partners, in our model single production dominates in parts of the parameter space. To maximize sensitivity to top partners, experiments should extend the menu of searches to encompass all available production and decay modes. Another interesting handle not used in the top partner searches so far is jet substructure: the top partner decay products, such as $t$, $Z$, $W$ and $h$, are typically relativistic in the lab frame in the relevant mass range, so that their hadronic decays can be identified as jets with unusual substructure. Recent phenomenological studies~\cite{Martin,boosted} show interesting potential of such searches.    

As a complementary handle, measurements of the Higgs and top properties at the LHC may be sensitive to deviations from the SM predicted by our model. While these predictions are quite model-dependent, our study indicates that large deviations in $h\to WW/ZZ$ and $h\to\gamma\gamma$ rates are possible. 

\vskip0.8cm
\noindent{\large \bf Acknowledgments} 
\vskip0.3cm
We are grateful to Monika Blanke, Yuval Grossman, Gala Nicolas Kaufman, Michael Saelim, and Javier Serra for useful discussions. J.B. and M.P. are supported by the U.S. National Science Foundation through grant PHY-0757868 and CAREER grant No. PHY-0844667. J.H. is  supported in part by the DOE under grant number DE-FG02-85ER40237. J.H. thanks Cornell University for hospitality during the course of this work. 
\begin{appendix}

\section{Masses, Mixing Angles and Couplings of the Top and Its Partner}

Ignoring the Goldstone fields that are eaten by the SM gauge bosons after EWSB, the sigma field $V$ has the form 
\beq
V = \exp (i a I) \left( \begin{tabular}{c} 0 \\ $f$ \end{tabular} \right)\,,
\eeq{sf}
where 
\beq
I \,=\, \left( \begin{tabular}{cc} $0$ & $1$ \\ $1$ & $0$ \end{tabular} \right)\,,
\eeq{Idef}
and
\beq
a = \frac{1}{\sqrt{2}}\frac{v+h}{f}\,.
\eeq{adef}
Here $v$ is the Higgs vev, and $h$ is the physical Higgs boson. The exponent can be easily expanded using the fact that $I^2=1$:
\beq
\exp (i a I) = \cos a + i \sin a \,I\,.
\eeq{sftrig}
The kinetic term of the sigma model has the form
\beq
{\cal L}_{\rm kin} = \left(D_\mu V\right)^\dagger \left(D_\mu V\right)\,,
\eeq{Lkin}
where $D_\mu$ is the covariant derivative. This term contains a canonically normalized kinetic term for the Higgs, as well as masses for the SM gauge bosons; in particular,
\beq
m_W^2 = \frac{1}{2} g^2 f^2 \sin^2 \bar{a}\,,
\eeq{mW}
where we defined $\bar{a} = v/(\sqrt{2}f)$. The measured value of $m_W$ can be used to compute $v$ from this formula; in the limit $f\to\infty$, $v$ tends to its SM value, 246 GeV. 

Using Eq.~\leqn{sftrig}, the top mass terms take the form
\beq
{\cal L}_{\rm mass} = (u_R^\dagger~U_R^\dagger)\, {\cal M} \,\left( \begin{tabular}{c} $u_L$ \\$U_L$ \end{tabular} \right) \,+\,~{\rm h.c.} \,, 
\eeq{tm1}
where
\beq
{\cal M} \,= \,f\,\left( \begin{tabular}{cc} $\lambda_1 \sin \bar{a}$ & $\lambda_1 \cos\bar{a}$ \\ $0$ & $\lambda_2$ \end{tabular} \right)\,.
\eeq{tm2}
Diagonalizing ${\cal M}^\dagger {\cal M}$, we find the masses of the top quark $t$ and its partner $T$:
\beq
m^2_{t,T} \,=\,\frac{(\lambda_1^2+\lambda_2^2)f^2}{2} \,\left(1\pm \sqrt{1-\frac{4\lambda_1^2\lambda_2^2\sin^2 \bar{a}}{(\lambda_1^2+\lambda_2^2)^2}}\right)\,.
\eeq{masses}
The rotation between gauge eigenstates $(u, U)$ and mass eigenstates $(t, T)$ is given by
\beqa
t_L &=& \cos\beta \,u_{L} - \sin\beta \,U_{L},~~~~~~~
T_{L} = \sin\beta \,u_{L} +\cos\beta \,U_{L}\,;\CR
t_R &=& \cos\alpha \,u_R - \sin\alpha \,U_{R},~~~~~~~
T_{R} = \sin\alpha \,u_R + \cos\alpha \,U_{R}\,,
\eeqa{rot2}
and the mixing angles are 
\beqa
\alpha &=& \half \tan^{-1} \frac{2\lambda_1\lambda_2 \cos\bar{a}}{\lambda_2^2-\lambda_1^2}\, ,\CR
\beta &=& \half \tan^{-1} \frac{\lambda_1^2 \sin2\bar{a}}{\lambda_2^2+\lambda_1^2\cos2\bar{a}}.
\eeqa{mixingsexact}
Mass and mixing angle formulas quoted in the main text are obtained by expanding in $v/f$ and keeping the leading order terms only.

It is also useful to invert these formulas and express the Lagrangian parameters $(\lambda_1, \lambda_2, f)$ in terms of physical parameters $(m_t, m_T, \alpha)$:
\beqa
f &=& \left( \frac{\sqrt{2}m_W}{g}\right)\,\frac{1}{x_t^{1/2}}\,\left(\cos^2\alpha + x_t \sin^2\alpha \right)^{1/2}\,\left(\sin^2\alpha + x_t \cos^2\alpha \right)^{1/2}\,,\CR
\lambda_1 &=& \left(\frac{gm_t}{\sqrt{2}m_W}\right) \, \frac{1}{\left(\cos^2\alpha + x_t\sin^2\alpha \right)^{1/2}} \,,\CR
\lambda_2 &=& \left(\frac{gm_t}{\sqrt{2}m_W}\right) \, \frac{1}{\left(\sin^2\alpha + x_t \cos^2\alpha  \right)^{1/2}}\,,
\eeqa{phystoL}
where $x_t=m_t^2/m_T^2$. For example, together with the second line of Eq.~\leqn{mixingsexact}, this expressions give 
the angle $\beta$ in terms of the physical parameters, which was used in the calculation of precision electroweak parameters in Sec.~\ref{sec:pew}: 
\beq
\sin\beta = \frac{x_t^{1/2}}{\sqrt{\cot^2\alpha + x_t}}\,.
\eeq{betaphys}

The couplings of the top and its partner to electroweak gauge bosons are given by 
\beqa
{\cal L}_{\rm g} &=& \frac{e}{\sqrt{2} s_w} \,b_L^\dagger \bar{\sigma}^\mu \left(\cos\beta t_L + \sin\beta T_L\right) W^-_\mu + {~\rm c.c.} \,\CR &+&
\left( g_{ttL} \,t_L^\dagger  \bar{\sigma}^\mu t_L + g_{ttR} \,t_R^\dagger  \bar{\sigma}^\mu t_R + g_{TTL} \,T_L^\dagger  \sigma^\mu T_L +g_{TTR} \,T_R^\dagger  \sigma^\mu T_R\right) Z_\mu \CR &+& g_{tTL} \,t_L^\dagger  \bar{\sigma}^\mu T_L Z_\mu+ {~\rm c.c.}\,,
\eeqa{Lg}
where
\beqa
g_{ttL} &=& \frac{e}{c_ws_w}\left(\frac{\cos^2\beta}{2}-\frac{2s_w^2}{3}\right);~~~g_{ttR} \,=\, - \frac{2es_w}{3c_w}\,;\CR
g_{TTL} &=& \frac{e}{c_ws_w}\left(\frac{\sin^2\beta}{2}-\frac{2s_w^2}{3}\right);~~~g_{TTR} \,=\, - \frac{2es_w}{3c_w}\,;\CR
g_{tTL} &=& \frac{e\sin2\beta}{4s_w c_w}\,.
\eeqa{Zcouplings}
Their couplings to the Higgs boson are 
\beq
{\cal L}_{\rm yuk} = - \left( C_{tth} t_L^\dagger t_R + C_{TTh} T_L^\dagger T_R + C^L_{Tth} t_R^\dagger T_L +
C^R_{Tth} T_R^\dagger t_L \right) \frac{h}{\sqrt{2}}\,+~{\rm h.c.},
\eeq{Lyuk}
where
\beqa
C_{tth} &=& \lambda_1 \,\cos\alpha\,\cos(\bar{a}-\beta);~~~
C_{TTh} = - \lambda_1\,\sin\alpha\,\sin(\bar{a}-\beta);\CR
C^L_{Tth} &=& -\lambda_1\,\cos\alpha \sin(\bar{a}-\beta);~~~
C^R_{Tth} = \lambda_1\,\sin\alpha \cos(\bar{a}-\beta).
\eeqa{h_couplings}
Finally, the Higgs boson coupling to the electroweak gauge bosons are given by 
\beq
{\cal L}_{hVV} = 2 \cos\bar{a} \left(m_W^2 W^{+\mu} W^-_\mu + \frac{1}{2} m_Z^2 Z^\mu Z_\mu \right) \frac{g h}{2 m_W} \,.
\eeq{hVV}
These couplings are suppressed compared to the SM values by a common factor, 
\beq
\cos\bar{a} = \sqrt{1-\frac{2m_W^2}{g^2f^2}}.
\eeq{shift}

\section{Loop Functions Appearing in Flavor Observables}

The $F$ functions that arise from calculating the box diagrams for $\Delta F = 2$ processes are given by~\cite{InamiLim}:~\begin{align}
F(x_i,x_j,M_W) = \frac{1}{(1-x_i) (1-x_j) } \left( 1 - \frac{7}{4} x_i x_j \right) + \frac{x_i^2 \log x_i}{(x_i-x_j) (1-x_i)^2} \left(1-2 x_j + \frac{x_i x_j}{4} \right) \nonumber \\
+ \frac{x_j^2 \log x_j}{(x_j-x_i) (1-x_j)^2} \left(1-2 x_i + \frac{x_i x_j}{4} \right) 
\end{align}
where the corresponding box diagrams have been calculated in Feynman-t'Hooft gauge.  Since we computed the mass eigenvalues for the top sector at all orders in the $v/f$ expansion, we also have the precise values for the $F$ functions. 

The loop functions appearing in the $b\to s\gamma$ amplitude are (see, for example, Ref.~\cite{SMbsgamma}):
\beqa
A^t_0(x) &=& \frac{-3x^3+2x^2}{2(x-1)^4} \log x + \frac{-22x^3+153x^2 -159x+46}{36(x-1)^3},\CR
F^t_0(x)&=& \frac{3x^2}{2(x-1)^4} \log x + \frac{-5x^3+9x^2 -30x+8}{12(x-1)^3}.
\eeqa{AandF}
The $B_s\to\mu^+\mu^-$ branching ratio is given by~\cite{Bmumu}
\beq
\frac{{\rm Br}(B_s\to\mu^+\mu^-)}{{\rm Br}(B_s\to\mu^+\mu^-)_{\rm SM}}\,=\,\left| 1+\frac{\bar{Y}}{Y_{\rm SM}}\right|^2,
\eeq{Bmm}
where
\beqa
& &\hskip-.7cm Y_{\rm SM} = \frac{x_t}{8}\left[ \frac{x_t-4}{x_t-1} +\frac{3x_t}{(x_t-1)^2}\log x_t\right],\CR
& &\hskip-.7cm \bar{Y} = s_\beta^2 \left[ \frac{2+2x_t-2x_t^2}{8(-1+x_t)} - \frac{x_t(2-x_t+2x_t^2)}{8(-1+x_t)^2}\log x_t + 
\frac{3+2x_t}{8}\log x_T + \frac{x_t}{8}\tan^2\alpha \right]. 
\eeqa{Ys}
Note that these expressions are only valid to order $(v/f)^2$.

\end{appendix}

\end{document}